\documentclass[%
 reprint,
 amsmath,amssymb,
 aps,
]{revtex4-2}

\usepackage[dvipsnames, svgnames, x11names]{xcolor}
\usepackage{graphicx}% Include figure files
\usepackage{dcolumn}% Align table columns on decimal point
\usepackage{bm}% bold math
\usepackage{amsmath}
\usepackage{multirow}
\usepackage{booktabs}
\usepackage{amsfonts}
\usepackage{makecell}
\usepackage{float}
\usepackage{diagbox}
\usepackage{tabularx}
\usepackage{float}
\definecolor{deepgreen}{rgb}{0, 0.5, 0}

\begin{document}

\preprint{APS/123-QED}

\title{Erasing Doppler Dephasing Error in Rydberg Quantum Gates }% Force line breaks with \\

\author{Rui Li$^{1}$}
\author{Jing Qian$^{2,3,4}$}
\email{Corresponding author: jqian1982@gmail.com}
\author{Weiping Zhang$^{1,4,5,6}$}

\affiliation{$^{1}$School of Physics and Astronomy, and Tsung-Dao Lee Institute, Shanghai Jiao Tong University, Shanghai, 200240, China}
\affiliation{$^{2}$State Key Laboratory of Precision Spectroscopy, Department of Physics, School of Physics and Electronic Science, East China Normal University, Shanghai, 200062, China
}
\affiliation{$^{3}$Chongqing Institute of East China Normal University, Chongqing, 401120, China}
\affiliation{$^{4}$Shanghai Research Center for Quantum Science, Shanghai, 201315, China}
\affiliation{$^{5}$Shanghai Branch, Hefei National Laboratory, Shanghai 201315, China}
\affiliation{$^{6}$Collaborative Innovation Center of Extreme Optics, Shanxi University, Taiyuan, Shanxi 030006, China}

%\affiliation{%
 %Authors' institution and/or address 
% This line break forced with %\textbackslash\textbackslash
%}%


\begin{abstract}
The Doppler dephasing error due to residual thermal motion of qubit atoms is a major cause of fidelity loss in neutral-atom quantum gates. Besides cooling and trapping advancements, few effective methods exist to mitigate this error.
In the present work, we introduce an error-erasing strategy that utilizes a pair of off-resonant fields to continuously dress the protected Rydberg state with an auxiliary state, which induces an opposite but enhanced sensitivity to the same source of Doppler dephasing error. Combining with an optimal control of laser pulses, we realize a family of Rydberg two-qubit controlled-NOT gates in Rb and Cs atoms that are fully robust to the Doppler dephasing error. We benchmark this gate operation with fidelity $F\approx0.9906$ at {\it any} temperature for a lower-excited auxiliary state, and a higher fidelity of $F\approx0.9965$ can be attained for a ground-state auxiliary state at a temperature of 50 $\mu$K. Our results significantly reduce atomic temperature requirements for high-fidelity quantum gates, and may provide fundamental guidance to practical error-tolerant quantum computing with neutral atoms.
\end{abstract}

\maketitle

\section{Introduction}

Errors restrict the fidelity of Rydberg gates in neutral-atom quantum computing and must be made sufficiently low by using versatile fault-tolerant techniques \textcolor{black}{\cite{Chao2018,PhysRevX.12.021049,PRXQuantum.3.010329,Wu2022,Ma2023,Scholl2023}}. So far the ability to reduce the sensitivity of gate operations to various intrinsic and technical errors by using robust pulses is a key capability for constructing a neutral-atom quantum computer \textcolor{black}{\cite{PhysRevA.80.013417,PhysRevResearch.4.033019}}, and the best reported fidelity for a two-qubit controlled-phase gate has reached 0.995 relying on optimal control strategy \textcolor{black}{\cite{Evered2023}}. Typically, a simple time-optimal gate with robust pulses maximizes the ideal gate fidelity in the absence of any error yet it can be substantially impacted if the type of errors occur \textcolor{black}{\cite{PhysRevResearch.5.033052,Chang_2023}}. Recent efforts have shown that gate protocols can be made natively robust to certain error sources either by modifying the cost function in optimization \textcolor{black}{\cite{PhysRevA.90.032329,PhysRevResearch.4.023155,PRXQuantum.4.020336,PhysRevLett.132.193801,Zhang_2024,PhysRevApplied.21.044012}} or by converting the feature of quasistatic errors \textcolor{black}{\cite{PRXQuantum.4.020335}}. However, to further improve the robustness of gates against certain type of errors beyond pure numerical methods, remains a great challenge to date.

As we know, decoherence from residual atomic motion fundamentally limits the gate fidelity in experiment \textcolor{black}{\cite{PhysRevA.105.042430,PhysRevLett.123.230501}}. Atoms, whether they are warm or cold, are not stationary inevitably leading to the motional dephasing due to inhomogeneous velocity distribution. Although a traditional two-photon transition with two counterpropagating excitation lasers can diminish the impact of Doppler dephasing nevertheless this improvement is very limited  in a sense \textcolor{black}{\cite{PhysRevLett.104.010502,PhysRevLett.104.010503,PhysRevLett.119.160502,PhysRevLett.123.170503,PhysRevLett.129.200501}} except for using more technically demanding three-photon excitation \textcolor{black}{\cite{PhysRevA.84.053409}}.
Therefore such a Doppler dephasing error remains a crucial resource of technical errors for promising applications in quantum information processing, and meanwhile is rarely mitigated unless actual colder temperatures $\sim \mu$K are reached \textcolor{black}{\cite{PhysRevX.2.041014,PhysRevLett.110.133001}}. 
With the recent demonstration of a coherence protection scheme in thermal atomic ensembles by Finkelstein {\it et.al.} \textcolor{black}{\cite{PhysRevX.11.011008}} where a collective state can be fully protected from inhomogeneous dephasing by employing off-resonant fields that dress it to an auxiliary sensor state, there is a significant interest in achieving a Doppler-error erased gate by applying this scheme.

In this work, we introduce a family of practical protocols for realizing two-qubit Rydberg controlled-NOT (CNOT) gates which exhibit full robustness to the Doppler dephasing error originating from thermal motion of qubit atoms. In combination with the optimal control method \textcolor{black}{\cite{PhysRevApplied.13.024059,Jandura2022timeoptimaltwothree,PhysRevA.109.022613}} our novel protocols employ a pair of off-resonant laser fields continuously dressing the protected Rydberg state with a lower-excited auxiliary state
that has an opposite but enhanced sensitivity to the same error source. The resulting gates have shown a full immunity to the Doppler dephasing error impacted on the Rydberg level at {\it any} temperature. All gate pulses including their amplitudes and phases as well as the gate duration are globally optimized avoiding the requirement for local addressing with Rydberg excitation lasers \textcolor{black}{\cite{PhysRevA.101.062309,PhysRevLett.131.170601}}.

To characterize the compatibility of our protocol to various atomic systems we explore the results using different dressing cases in Rb and Cs atoms and show the choice of a higher sensitivity factor is key for improving the gate performance. Moreover we also discuss the method to suppress spontaneous decay errors from the auxiliarily excited state which ultimately limits the gate fidelity presently \textcolor{black}{\cite{PhysRevApplied.10.034006}}. Utilizing a ground-state auxiliary state, although the incoherent decay error can be avoided we find that the gate turns to be slightly sensitive
 to Doppler shifts owing to the existence of an unprotected Rydberg state. Finally, with the increase of finite temperature of qubit atoms, we confirm that the newly-proposed dressing protocols are significantly superior to the typical no-dressing gates, by mutually having a higher fidelity and a lower Doppler dephasing error over a very wide range of atomic temperatures.

\section{Model and Hamiltonians}

\begin{table*}
\caption{Optimized gate parameters for three laser amplitudes $\Omega_r(t),\Omega^{\prime}_r(t),\Omega_d$(in unit of $2\pi\times$MHz), laser phases $\phi(t),\phi^\prime(t)$(in unit of $2\pi$) as well as the gate duration $T_g$(in unit of $\mu$s). In all cases the dressing-field detuning is $\Delta_d\approx\Omega_d/0.698$ ensured by the insensitive condition. The blockade strength is \textcolor{black}{$V/2\pi =  200$ MHz}. The last column presents the ideal gate fidelity in the absence of any decay ($\gamma_{r}=\gamma_a=0$).  }
\begin{ruledtabular}
\renewcommand{\arraystretch}{1.1}
\setlength{\tabcolsep}{0mm}{
\begin{tabular}{cccccccccccc}
Case & $\quad$ & \multicolumn{1}{c}{$T_g$} & \multicolumn{2}{c}{$\Omega_r(t)$} & \multicolumn{2}{c}{$\Omega^{\prime}_r(t)$}  & \multicolumn{3}{c}{$\phi(t),\phi^\prime(t)$} &  $\Omega_d$ &  $F$(ideal)\\
\hline
$\quad$ & $\quad$ & $\quad$ & $\Omega_{\max}$ & $\omega$ & $\Omega^{\prime}_{\max}$ & $\omega^{\prime}$ & $\delta_0$ & $\delta_1$ & $\delta_2$  &$\quad$ &$\quad$\\
\hline
\multirow{2}*{no-dressing} & linear & 1.00 & 9.87 & 0.1946 & 10.0 & 0.1938 & 4.90 & $\times$ & $\times$ & $\times$  & \textcolor{black}{0.99945}\\
~ & composite & 0.62 & 9.19 & 0.1018 & 8.96 & 0.1026 & -0.117 & 0.589 & -0.0006  & $\times$ & \textcolor{black}{0.99959}\\
\hline
\multirow{2}*{with-dressing} & linear & 3.18 & 9.56 & 0.1007 & 9.59 & 0.1007 & -4.97 & $\times$ & $\times$ & 195.7  & \textcolor{black}{0.99726}\\
~ & composite & 3.60 & 9.89 & 0.1091 & 9.95 & 0.1093 & -4.77 & -0.57 & -2.07 & 201.4  & \textcolor{black}{0.99971}\\
\end{tabular}
}
\end{ruledtabular}
\label{table1}
\end{table*}

\begin{figure}
\centering
\includegraphics[width=0.45\textwidth]{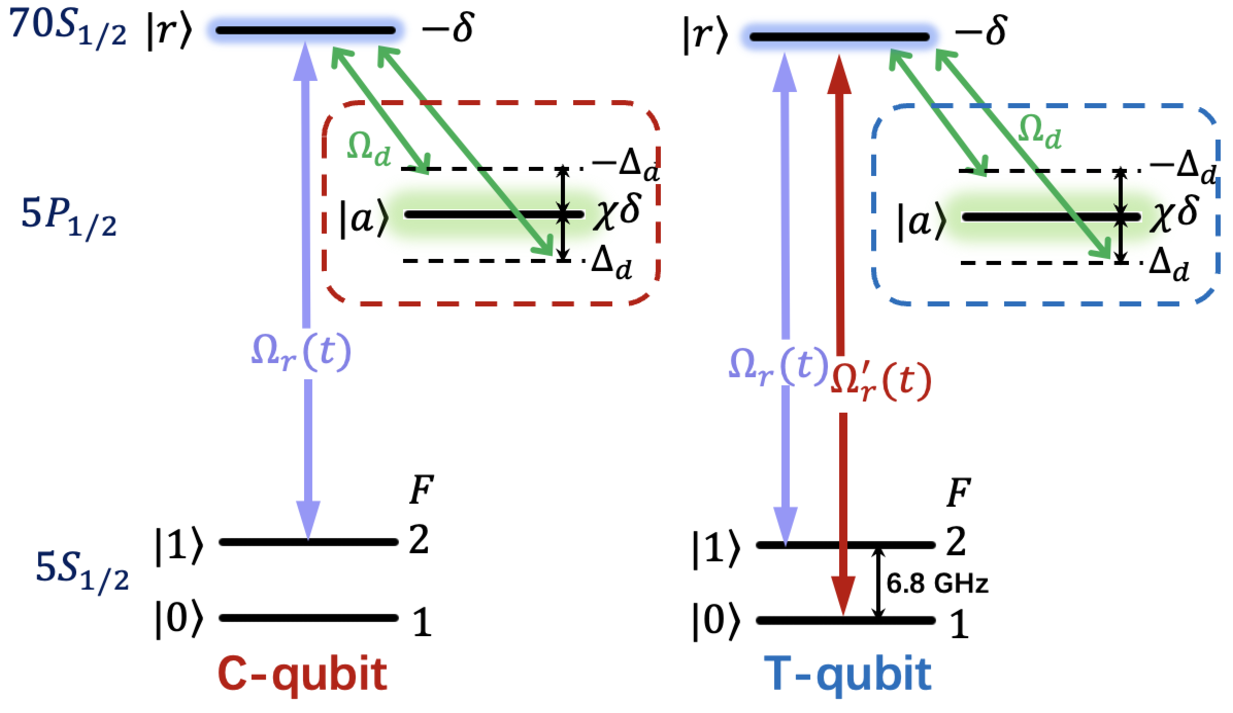} 
\caption{Doppler dephasing error erased two-qubit CNOT gates. Level scheme: 
qubits $|0\rangle,|1\rangle$ are encoded in the hyperfine clock states \textcolor{black}{$|5S_{1/2}, F=1,m_F = 0\rangle$ and $|5S_{1/2}, F=2,m_F = 0\rangle$} of two $^{87}$Rb atoms. The $|0\rangle$ and $|1\rangle$ states are coupled to the Rydberg state $|r\rangle$ with two-photon Rabi frequency of real amplitudes $\Omega_r,\Omega_r^{\prime}$ and phases $\phi,\phi^{\prime}$, which undergoes an unknown Doppler shift $-\delta$ due to the atomic velocity. Additionally, two dressing fields with same Rabi frequency $\Omega_d$ and opposite detunings $\Delta_d,-\Delta_d$ enable the coupling between $|r\rangle$ and an auxiliary state $|a\rangle$ ($|a\rangle=|5P_{1/2}\rangle$ is a lower excited state). The overall transition frequency between $|1\rangle$ and $
|a\rangle$ will experience an energy shift $\chi\delta$, possibly opposite and larger as required by the insensitive condition (see Appendix \ref{appendixa}). The most intrinsic error treating as the ultimate limit for the gate fidelity is the spontaneous decay from state $|a\rangle$ (see Sec. \ref{limitation}). }
\label{Fig.model}
\end{figure}

% {\it Level scheme description.—}

To implement the protection for a Doppler dephasing-error erased gate we assume the level scheme as shown in Fig. \ref{Fig.model}. Each atom is modeled as a four-level system with long-lived ground states $|0\rangle,|1\rangle$, an uppermost Rydberg state $|r\rangle$ with lifetime $1/\tau_r$ and an auxiliarily intermediate state $|a\rangle$ with lifetime $1/\tau_a$. The traditional coupling between $|1\rangle$ and $|r\rangle$ is enabled by a global two-photon laser pulse with time-dependent Rabi frequency $\Omega_r(t)$, between $|0\rangle$ and $|r\rangle$ by $\Omega_r^\prime(t) $. Moreover, the laser frequency is tuned to be resonant with the transition between $|1\rangle$(or $|0\rangle$) and $|r\rangle$. Additionally, we require a pair of dressing fields off-resonantly coupling $|r\rangle$ and $|a\rangle$ with a same strength $\Omega_d$ and opposite detunings $\Delta_d,-\Delta_d$, generated by using an electro-optic modulator \textcolor{black}{\cite{PhysRevX.11.011008}}. As a consequence, when the Doppler effect causes an uncertain detuning error of $|1\rangle\to|r\rangle$ transition, i.e. $-\delta = \vec{k}_r\cdot\vec{v}$, state $|a\rangle$ can also be affected by the same source of inhomogeneity from same atomic velocity, experiencing an unknown energy shift $\chi\delta = (\vec{k}_r+\vec{k}_a)\cdot\vec{v}$ with $\chi=|1+\vec{k}_a/\vec{k}_r|$. Here the opposite sign ($-\delta,\chi\delta$) is ensured by the choice of different wavevectors of lasers $(\vec{k}_a,\vec{k}_r)$ as well as their propagation directions. For simplicity the two-photon transition on the target qubit driving $|0\rangle$ and $|r\rangle$ is enabled by a same wavevector $\vec{k}_r$.

We require the sensitivity factor to be $\chi\geq1$, which means the dressing state $|a\rangle$ senses the fluctuation at least as comparable as the Rydberg state $|r\rangle$. This process can be achieved by an optical transition whose wavevector $\vec{k}_a$ has an opposite direction and a much larger magnitude (i.e. a shorter wavelength). In two-photon transitions the excitation wavevector $\vec{k}_r$ is a vectorial sum of participating fields' wavevectors. Their intermediate state $|e\rangle$ (not shown, we choose $|e\rangle=|5P_{3/2}\rangle$) and dressing state $|a\rangle$ determines the ratio $\vec{k}_a/\vec{k}_r$. Note that the transition $|r\rangle\to |a\rangle$ should have an enhanced and opposite velocity sensitivity as compared to that of $|1\rangle\to|r\rangle$, consequently the choice of $|a\rangle$ is vital. A nearby long-lived Rydberg state is impossible to be $|a\rangle$. Thereby we first suggest protecting the Rydberg state $|r\rangle=|70S_{1/2}\rangle$ by dressing it to a lower excited state $|a\rangle = |5P_{1/2}\rangle$ in $^{87}$Rb atoms. This choice provides $\vec{k}_r=2\pi(\lambda_{480}^{-1}-\lambda_{780}^{-1})= 5.035$ $\mu$$m^{-1}$, $\vec{k}_a = -2\pi/\lambda_{475}=-13.228$ $\mu$$m^{-1}$, arising the sensitivity factor $\chi \approx 1.627$ if $\Omega_d/\Delta_d \approx 0.698$. 
In Sec. VI we alternatively introduce $|a\rangle$ to be a hyperfine ground state where the one-photon transition between Rydberg and auxiliary states must be performed by an ultraviolet laser \textcolor{black}{\cite{Thoumany:09}}.

The total Hamiltonian governing the dynamics of two atoms reads
\begin{equation}
    \mathcal{H} = \mathcal{H}_c \otimes I+I \otimes \mathcal{H}_t + V|rr\rangle\langle rr|
\end{equation}
with single-qubit Hamiltonians
\begin{eqnarray}
    \mathcal{H}_c&=&\frac{\Omega_r(t)}{2}|1\rangle\langle r|+ \frac{\Omega_d(e^{i\Delta_dt} + e^{-i\Delta_dt})}{2}|r\rangle\langle a|+\text{H.c.} \nonumber\\
    &-&\delta|r\rangle\langle r|+\chi\delta|a\rangle\langle a|  \nonumber\\
   \mathcal{H}_t&=&\frac{\Omega_r(t)}{2}|1\rangle\langle r|+\frac{\Omega^{'}_r(t)}{2}|0\rangle\langle r|+\frac{\Omega_d(e^{i\Delta_dt} + e^{-i\Delta_dt})}{2}|r\rangle\langle a|\nonumber\\
   &+&\text{H.c.} -\delta| r \rangle\langle r|+\chi\delta|a\rangle\langle a| \nonumber
\end{eqnarray}
and $I$ the identity operator.
Here, $V$ denotes the strength of van der Waals interaction for Rydberg pair state $|rr\rangle$, and $\Omega_r(t)= |\Omega_r(t)|e^{i\phi(t)}$, $\Omega_r^\prime(t)= |\Omega_r^\prime(t)| e^{i\phi^\prime(t)}$ are the laser Rabi frequencies.  We stress a global drive of both laser fields, so as to eliminate the constraint of local operations. Besides two atoms are placed at short distances such that the interaction strength $V$ is much larger than laser Rabi frequencies (i.e. $V\gg |\Omega_r(t)|,|\Omega_r^\prime(t)|$), thus resulting in a strong suppression of simultaneous excitation of both atoms to the Rydberg state \textcolor{black}{\cite{Urban:09,Gaetan2009}}.
In the following we are interested in finding robust pulses $\Omega_r(t),\Omega_r^\prime(t)$ that are insensitive to the variation of Doppler dephasing error $\delta$.

\section{Doppler dephasing robust pulses}
\label{rob}

We proceed by addressing the suppression of Doppler error via the combination of gate protocols relying on optimal control method \textcolor{black}{\cite{PhysRevA.84.042315,Muller_2022,PhysRevA.109.062603}}. A two-qubit CNOT gate can be realized by using numerically optimized continuous pulses $\Omega_r,\Omega_r^\prime,\Omega_d$ with series of tunable parameters. To implement a time-optimal gate that operates within a duration of $T_g$, if
for the no-dressing case as discussed in our prior work \textcolor{black}{\cite{PhysRevApplied.17.024014}}, states $|00\rangle$ and $|01\rangle$ experience single-qubit rotations with two resonant couplings $\Omega_r(t),\Omega_r^\prime(t)$, requiring $|00\rangle\to |00\rangle$ and $|01\rangle\to |01\rangle$ at $t=T_g$. While states $|10\rangle$ and $|11\rangle$ are off-resonantly coupled due to the blockade constraint
and global drive. We expect the exact conversion of $|10\rangle\to |11\rangle$ and $|11\rangle\to |10\rangle$ for a CNOT gate with optimal parameters. Finally we can measure the fidelity of gate with a commonly used definition as
\begin{equation}
    F = \frac{1}{4}\text{Tr}\left[\sqrt{\sqrt{\mathcal{O}}\rho(t=T_g)\sqrt{\mathcal{O}}}\right]
\end{equation}
having considered the average effect of four computational basis states $\{|00\rangle,|01\rangle,|10\rangle,|11\rangle\}$, where $\mathcal{O}$ is the ideal unitary matrix for
the gate and $\rho(t=T_g)$ is the realistic output density matrix at the end of gate operation. To solve the dynamics of each basis state we use the Liouville-von Neumann equation with Lindblad relaxation terms \textcolor{black}{\cite{PhysRevA.95.012708}}
\begin{equation}
  \partial \rho(t)/\partial t = -i[\mathcal{H},\rho]+\mathcal{L}_r[\rho]+\mathcal{L}_a[\rho]  \label{rro}
\end{equation}
where 
\begin{eqnarray}
    \mathcal{L}_r[\rho] &=& \sum_{i\in\{0,1,a\}}(L_{ir}\rho L_{ir}^\dagger-\frac{1}{2}[L_{ir}^\dagger L_{ir}\rho + \rho L_{ir}^\dagger L_{ir}]) \nonumber \\
    \mathcal{L}_a[\rho] &=& \sum_{j\in\{0,1\}}(L_{ja}\rho L_{ja}^\dagger-\frac{1}{2}[L_{ja}^\dagger L_{ja}\rho + \rho L_{ja}^\dagger L_{ja}]) \nonumber
\end{eqnarray}
and $L_{ir}=\sqrt{\gamma_r/3}|i\rangle\langle r|$, $L_{ja}=\sqrt{\gamma_a/2}|j\rangle\langle a|$ stands for Lindblad operators that correspond to the spontaneous emission from each excited state to each ground state.

In the presence of an auxiliary state $|a\rangle$ (with-dressing case) optimization of pulses for the gate requires more tunable parameters, leading to the entire evolution difficult to satisfy. In this work, in order to ease the experimental implementation we use smooth modulation for both laser amplitude $|\Omega_r^{(\prime)}(t)|$ and phase $\phi^{(\prime)}(t)$, and set the dressing field $\Omega_d$ as a constant value to be optimized. Specifically we perform a smooth phase modulation whose slope also represents the two-photon detuning value \textcolor{black}{\cite{Evered2023,PhysRevApplied.16.064040}}. Two kinds of robust pulses are considered now: one with a phase profile given by a linear function
\begin{equation}
    \phi(t) = \delta_0 t
\end{equation}
corresponding to a fixed two-photon detuning $\delta_0$ to $|r\rangle$, 
and a second one with a composite phase profile
\begin{equation}
    \phi(t) = \delta_0 t + \delta_1\sin(\frac{4\pi t}{T_g}) + \delta_2\cos(\frac{2\pi t}{T_g})
\end{equation}
which results in a more sufficient modulation for the two-photon detuning. We assume $\phi^\prime = \phi$ for simplicity. Except for the varying phase, the varying amplitude depends on an assumption of smooth Gaussian profiles
\begin{equation}
    |\Omega_r(t)|=\Omega_{\max}e^{-\frac{(t-T_g/2)^2}{2\omega^2}}, |\Omega^{'}_r(t)|=\Omega^{'}_{\max}e^{-\frac{(t-T_g/2)^2}{2\omega^{\prime 2}}}
\end{equation}
with $\Omega_{\max},\Omega_{\max}^\prime$ the maximal amplitudes and $\omega,\omega^\prime$ the pulse widths.
Following Ref.\textcolor{black}{\cite{Li_2023}} we choose the numerical genetic algorithm with single target of maximizing the gate fidelity $F$ by globally optimizing all pulse parameters. Note that the intrinsic decay error from finite lifetimes is also considered by numerical optimization ensuring a minimal time-spent on intermediate excited states $|r\rangle$ and $|a\rangle$ \textcolor{black}{\cite{PhysRevResearch.4.033019}}. In addition we add the gate duration $T_g$ as another quantity to be optimized which is limited by the maximal amplitudes. For constituting a practical gate that can reduce the spontaneous decay from excited states,
we choose a higher limitation for the two-photon Rabi frequency $(\Omega_{\max},\Omega_{\max}^\prime)/2\pi\leq 10$ MHz which enables the gate duration $T_g\leq 1.0$ $\mu$s in the no-dressing case \textcolor{black}{\cite{PhysRevApplied.14.054058}}. Whilst this duration scale will be much prolonged in the dressing case due to the extra population transfer between $|r\rangle$ and $|a\rangle$. 

\begin{figure}
\centering
\includegraphics[width=0.49\textwidth]{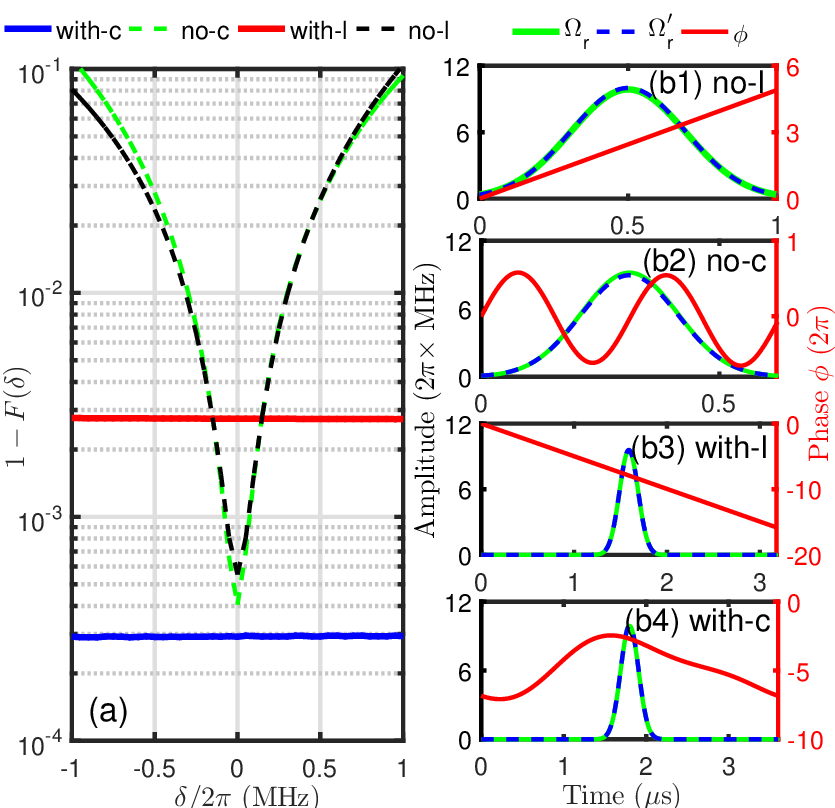} 
\caption{(a) Dependence of the gate infidelity $1-F(\delta)$ on the Doppler shift $\delta$ in the absence of decay. The no-dressing case with linear (or composite) phase modulation is denoted by {\it no-l} (or {\it no-c}). Similarly, for the dressing case, results are denoted by {\it with-l} and {\it with-c}. (b1-b4)
Laser amplitudes and phase profiles with respect to the four cases in (a).}
\label{Fig.insensitivity}
\end{figure}

The optimized gate parameters in no-dressing and with-dressing cases are presented in Table \ref{table1} in which two different phase profiles (linear and composite) are applied. The amplitude and phase profiles are plotted in Fig.\ref{Fig.insensitivity}(b1-b4) and the corresponding infidelities $1-F(\delta)$ as a function of $\delta$ are shown in Fig.\ref{Fig.insensitivity}(a). It is explicit that the no-dressing case has no robustness to the Doppler error $\delta$ no matter how to modulate the laser phase profile. Thereby once $|\delta|$ grows the infidelity has an exponential increase \textcolor{black}{\cite{PRXQuantum.4.020336}}, quickly reaching as high as $\sim0.1$ at $|\delta|/2\pi=1.0$ MHz. To quantify the performance of dressing case, Fig.\ref{Fig.insensitivity}(a) also presents the infidelity under the help of dressing fields by performing same optimization of all gate parameters. We find the ideal infidelity in both dressing cases stays constant for any $\delta$ implying a perfect protection from the Doppler dephasing error. 
It is more interesting that, by implementing a sufficiently composite modulation of the pulse phase the infidelity can even outperform the linear modulation by one order of magnitude staying at $\sim 3.0\times 10^{-4}$. 

%which also denotes the number of constant Doppler dephasing error in the scheme.

For completeness we also calculate the time-integrated intermediate-state population. The total time-spent in Rydberg state $|r\rangle$ of the no-dressing case is quite large $P_{r}\approx (0.8372,0.6536)$ $\mu$s corresponding to the linear and composite profiles which is mainly contributed by a wide laser-amplitude modulation over the entire gate execution [see (b1-b2)]. However, for the protection mechanism with dressing fields the real pulse widths $\omega,\omega^\prime$ are strongly narrowed although at the cost of a prolonged gate time [see (b3-b4)], which achieves a much shorter time-spent in Rydberg $P_r\approx(0.1133,0.0742)$ $\mu$s and auxiliary $P_{a}\approx (0.0705,0.0466)$ $\mu$s states, as compared to the no-dressing case. From comparison, we emphasize that, although the average time-spent in the auxiliary state $|a\rangle$ has been deeply minimized by optimization during the dressing case, our gate still suffers from a big decay error. Because the low-lying excited state $|a\rangle$ ($|5P_{1/2}\rangle$) is short-lived with its decay rate $\gamma_a$ typically 3 orders of magnitudes larger than $\gamma_r$ of state $|r\rangle$, 
which ultimately limits the realistic gate fidelity. \textcolor{black}{The discussion for this dominant decay error will be presented in Sec. \ref{limitation}}.

\section{Application to other dressing cases} 

\begin{figure}
\centering
\includegraphics[width=0.36\textwidth]{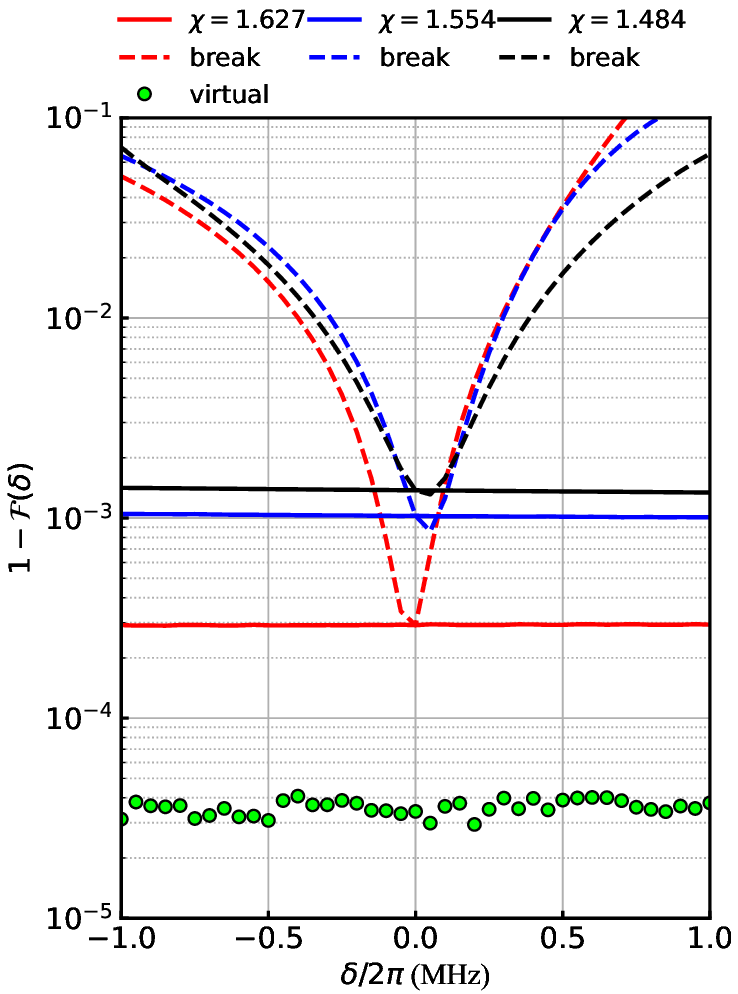} 
\caption{The gate infidelity $1-F(\delta) $ for different sensitivity factors $\chi = (1.627,1.484,1.554)$ denoted by solid lines. The corresponding dashed lines present the breakdown of the insensitive condition by increasing $\chi$ to 15 while all optimized parameters are kept unvaried. In addition, a virtually optimized case with a large $\chi=15$ is shown by full circles. No decay is included in the calculation. }
\label{Fig.break}
\end{figure}

\begin{table}
\caption{Choice of different intermediate states $|e\rangle$ and auxiliary states $|a\rangle$ for two-qubit CNOT gates in $^{87}$Rb and $^{133}$Cs atoms. The effective two-photon wavevector is $\vec{k}_r = 2\pi(\lambda^{-1}_{\rm{up}}-\lambda^{-1}_{\rm{lower}})$ in which $\lambda_{\rm{up}}$(nm) and $\lambda_{\rm{lower}}$(nm) represent the laser wavelengths of upper and lower transitions respectively, and $\vec{k}_a=-2\pi/\lambda_a$ with $\lambda_a$(nm) the wavelength of the optical dressing fields. $\tau_a$($\mu$s) denotes the lifetime of state $|a\rangle$. $\vec{k}_r$ and $\vec{k}_a$ are shown with a unit of $\mu$m$^{-1}$. We highlight the selected three cases in bold in which case (c) has been studied in Sec. \ref{rob}.}
\begin{ruledtabular}
\renewcommand{\arraystretch}{1.4}
\setlength{\tabcolsep}{0.2mm}{
\begin{tabular}{c|cc|cc|cc|cc}
& \multicolumn{4}{c|}{$^{87}$Rb} & \multicolumn{4}{c}{$^{133}$Cs} \\
\hline
Case & \multicolumn{2}{c|}{(a,b)} & \multicolumn{2}{c|}{(c,d)} & \multicolumn{2}{c|}{(e,f)} & \multicolumn{2}{c}{(g,h)}\\
\hline
$|e\rangle$ & \multicolumn{2}{c|}{$5P_{1/2}$} & \multicolumn{2}{c|}{$5P_{3/2}$} & \multicolumn{2}{c|}{$6P_{1/2}$} & \multicolumn{2}{c}{$6P_{3/2}$}\\
$\lambda_{\rm{up}}$ & \multicolumn{2}{c|}{475} & \multicolumn{2}{c|}{480} & \multicolumn{2}{c|}{495} & \multicolumn{2}{c}{509} \\
$\lambda_{\rm{lower}}$ & \multicolumn{2}{c|}{795} & \multicolumn{2}{c|}{780} & \multicolumn{2}{c|}{895} & \multicolumn{2}{c}{852} \\
$\vec{k}_r$ & \multicolumn{2}{c|}{5.324} & \multicolumn{2}{c|}{5.035} & \multicolumn{2}{c|}{5.673} & \multicolumn{2}{c}{4.969} \\
\hline
$|a\rangle$ & $5P_{1/2}$ & $5P_{3/2}$ & $5P_{1/2}$ & $5P_{3/2}$ & $6P_{1/2}$ & $6P_{3/2}$ & $6P_{1/2}$ & $6P_{3/2}$\\
$\lambda_a$ & 475 & 480 & 475 & 480 & 495 & 509 & 495 & 509\\
$|\vec{k}_a|$ & 13.228 & 13.089 & 13.228 & 13.089 & 12.693 & 12.344 & 12.693 & 12.344 \\
\hline
$\tau_a$ & 0.158 & 0.150 & 0.158 & 0.150 & 0.200 & 0.174 & 0.200 & 0.174\\
\hline
$\chi$ &\textbf{1.484} & 1.458 & \textbf{1.627} & 1.6  & 1.238 & 1.176 & \textbf{1.554} & 1.484 
\end{tabular}
}
\end{ruledtabular}
\label{table2}
\end{table}

\begin{table*}
\caption{Optimized gate parameters based on the choice of different sensitivity factors $\chi=(1.627,1.484,1.554)$ in Rb and Cs source atoms, corresponding to dressing levels in Cases (c), (a) and (g) of Table \ref{table2}. Besides the optimized parameters, the required dressing-field detuning $\Delta_d$, the average time-spent in Rydberg state $P_r$ and in auxiliary state $P_a$, the ideal gate fidelity $F$ are also given in the last four columns. We highlight the last row in green suggesting an ideally large sensitivity factor $\chi = 15$ regardless of the realistic atomic energy levels. This choice can provide a pronounced improvement in the ideal gate fidelity because of the far off-resonant coupling ($\Delta_d\gg\Omega_d$) to the auxiliary state $|a\rangle$. }
\begin{ruledtabular}
\renewcommand{\arraystretch}{1.1}
\setlength{\tabcolsep}{1.5mm}{
\begin{tabular}{ccccccccccc|cccc}
Atom & $\chi$ & \multicolumn{1}{c}{$T_g$} & \multicolumn{2}{c}{$\Omega_r(t)$} & \multicolumn{2}{c}{$\Omega^{\prime}_r(t)$}  & \multicolumn{3}{c}{$\phi(t),\phi^\prime(t)$} &  $\Omega_d$ & $\Delta_d$ & $P_r$ & $P_a$  & $F$(ideal)\\
\hline
$\quad$ & $\quad$ & $\quad$ & $\Omega_{\max}$ & $\omega$ & $\Omega^{\prime}_{\max}$ & $\omega^{\prime}$ & $\delta_0$ & $\delta_1$ & $\delta_2$  &$\quad$ &$\quad$  & $\quad$ & $\quad$ & $\quad$\\
\hline
\multirow{2}*{Rb} & 1.627 & 3.60 & 9.89 & 0.1091 & 9.95 & 0.1093 & -4.77 & -0.57 & -2.07 & 201.4  & 288.5 & 0.0742 & 0.0466  & 0.99971\\
~ & 1.484 & 3.59 & 9.70 & 0.1086 & 9.70 & 0.1080 & -15.0 & 2.72 & 0.874  & 262.4 & 362.0 & 0.0896 & 0.0612  & 0.99880\\
\hline
Cs & 1.554 & 3.55 & 9.43 & 0.1073 & 9.47 & 0.1075 & -7.37 & 0.54 & -0.98 & 218.6  & 307.3 & 0.0977 & 0.0638  & 0.99897\\
\hline
\textcolor{deepgreen}{Virtual} & \textcolor{deepgreen}{15} & \textcolor{deepgreen}{2.12} & \textcolor{deepgreen}{10.00} & \textcolor{deepgreen}{0.2383} & \textcolor{deepgreen}{9.12} & \textcolor{deepgreen}{0.2573} & \textcolor{deepgreen}{10.00} & \textcolor{deepgreen}{-1.09} & \textcolor{deepgreen}{-0.15}  & \textcolor{deepgreen}{240.0} & \textcolor{deepgreen}{945.4} & \textcolor{deepgreen}{0.2837} & \textcolor{deepgreen}{0.0192}  & \textcolor{deepgreen}{0.99997}\\
\end{tabular}
}
\end{ruledtabular}
\label{table3}
\end{table*}

Doppler dephasing induced by the residual thermal motion of qubit atoms occurs across various atomic systems which dominantly limits the gate performance in the field of quantum computing \textcolor{black}{\cite{PhysRevA.91.012337,PhysRevApplied.13.024008}}. In Sec. \ref{rob} we have used the atomic parameters corresponding to a Rb source atom with $|e\rangle=|5P_{3/2}\rangle$ and $|a\rangle = |5P_{1/2}\rangle$ arising the sensitivity factor $\chi = 1.627$ ($\Omega_d/\Delta_d \approx 0.698$). In order to verify the generality of our scheme adapted for more atomic systems we illustrate other double-dressing examples in Rb and Cs atoms in Table \ref{table2}. Note that the sensitivity factor $\chi$ is crucial that determines the degree of off-resonant dressing between $|r\rangle$ and $|a\rangle$, consequently affecting the time-spent in these states. To analyze the role of $\chi$ we newly optimize all gate parameters for cases (a) and (g) of Table \ref{table2}, as compared to the original case (c). As shown in Fig.\ref{Fig.break}, we find that cases (a) and (g) remain a steady sensitivity to the variation of Doppler shift $\delta$ which confirms the existence of an efficient protection mechanism from state $|a\rangle$. However, once the insensitive condition breaks, the Doppler shift on $|r\rangle$
can no longer be perfectly protected arising an expectedly exponential enhancement of the gate infidelity (see Fig.\ref{Fig.break}, dashed lines).
In Table \ref{table3} we give the three sets of optimized parameters in Rb and Cs atoms. For comparison we further introduce another set of virtually optimized parameters with a large $\chi=15$ which results in a far off-resonant coupling {i.e.} 
$\Delta_d\gg\Omega_d$ ensuring a much shorter time-spent in $|a\rangle$. Numerical results based on the virtual case are displayed in Fig.\ref{Fig.break} with full circles. It is obvious that, this case is also fully robust to the Doppler dephasing error and meanwhile benefits from a lower infidelity $1-F\sim 3\times 10^{-5}$ due to the deep minimization of time-spent in state $|a\rangle$ at off-resonant condition.

Finally we emphasize that the extension of this Doppler error erased protocol to more atomic sources (Rb, Cs) is possible as long as a suitably dressing state $|a\rangle$ featuring 
an opposite but enhanced velocity sensitivity for the $|r\rangle\to |a\rangle$ transition, 
can be found. Furthermore we note that the ideal gate fidelity can even be improved to be $>0.9999$ (last column in Table \ref{table3}) if the insensitivity factor $\chi$ grows. This is interesting in the limit where a 
long-lived auxiliary state is dressed such that the intrinsic decay error can be entirely avoided. \textcolor{black}{In order to diminish the dissipative effect from the auxiliary state, in Sec. \ref{prospective}, we introduce an improved protocol by dressing Rydberg state to a hyperfine ground state (not a lower excited state) through an one-photon transition. The central idea of protocol is to avoid the large spontaneous decay from $|5P_{1/2}\rangle$ by replacing with a more stable state.}

\section{Realistic gate performance with robust pulses}
\label{realistic}

\begin{figure}
    \centering
    \includegraphics[width=0.48\textwidth]{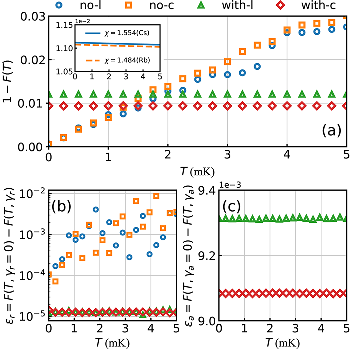}
    \caption{(a) The gate infidelity $1-F(T)$ at different values of atomic temperature where the decays of Rydberg and dressing states are both included, corresponding to the four cases described in Table \ref{table1}. The inset of (a) shows infidelities at the cases of $\chi = 1.484$ and $\chi = 1.554$ (see Table \ref{table3}). (b-c) Simulation of the realistic decay errors $\varepsilon_r$(for Rydberg state $|r\rangle$ at $\gamma_a = 0$) and $\varepsilon_a$(for intermediate state $|a\rangle$ at $\gamma_r = 0$) under different $T$. For a given $T$, each point denotes an average over 300 random realizations. Here the specific decay rates considered are $\gamma_r =1/\tau_r= 2.6$ kHz, $\gamma_a=1/\tau_a =2\pi\times 1.0$ MHz.}
    \label{fig:temperature}
\end{figure}

\subsection{Gate performance}

To gain insight on how Doppler dephasing error affects the realistic gate performance with robust pulses we now include the spontaneous decays of Rydberg and dressing states at a finite temperature $T$. During the gate execution, both the ground-Rydberg and the ground-dressing transitions will inevitably suffer from atomic motional dephasing, which make the real laser frequency perceived by the moving qubit atoms deviate from its ideal value \textcolor{black}{\cite{PhysRevA.97.053803}}. 
This can be estimated as the level detuning changes $\delta$ and $\chi\delta$, respectively for Rydberg and auxiliary states. Although the actual Doppler shifts of two atoms are different due to the uncorrelated atomic velocities, our gate protocol suggests an independent protection for each Rydberg state with an individual dressing state so the anti-symmetric detuning error does not work here \textcolor{black}{\cite{PRXQuantum.4.020336}}.
The velocity $\vec{v}$ originates from the finite temperature $T$ of trapped atoms constituting an intrinsic source of randomness. Here 
we assume $\vec{v}$ is randomly drawn from an one-dimensional Gaussian distribution with width $v_{rms} = \sqrt{k_B T/m}$, where $k_B$ is the Boltzmann constant and $m$ is the mass of atoms. This choice arises the detuning changes $\delta = \vec{k}_r\cdot \vec{v}$ and $\chi\delta = (\vec{k}_r+\vec{k}_a)\cdot\vec{v}$ are also random values. Then we calculate the gate infidelity caused by residual thermal motion of atoms, represented by the relationship between $1-F(T)$ and the atomic temperature $T$.

Numerical results are summarized in Fig.\ref{fig:temperature}(a). We first focus on two no-dressing cases. As expected the gate infidelity dramatically increases as $T$ grows no matter how to modulate the phase profile \textcolor{black}{\cite{PhysRevApplied.18.044042}}. Because without the use of protection mechanism the gate protocol is ideally optimized in the absence of Doppler error, i.e. at $T=0$. As $T$ increases both the Rydberg decay error $\varepsilon_r$ and Doppler dephasing error are dominant. At $T= 5$ mK we observe that these two errors contribute at a same level $\sim 10^{-2}$(also see Fig.\ref{fig:temperature}b), leading to the total infidelity as high as $1-F \approx 0.03$. In contrast, for the newly-proposed with-dressing case the infidelity of both cases ({\it with-l} and {\it with-c}) perfectly preserves a constant for {\it any} temperature, which means the Doppler dephasing error has been truly erased that no longer depends on the atomic temperature. Finally we note that, the selected choices ($\chi = 1.627,1.484,1.554$, see inset of Fig.\ref{fig:temperature}a) for Rb and Cs atoms can all contribute a Doppler-error erased gate with the increasing of atomic temperatures.

\subsection{Major limitation}
\label{limitation}

 Despite the perfect insensitivity to the Doppler dephasing error we have to admit a major limitation existing in the gate protocol, originating from the choice of a short-lived dressing state $|a\rangle$ \textcolor{black}{\cite{PhysRevLett.110.103001,doi:10.1126/science.1217901}}. For the chosen excited state $|a\rangle=|5P_{1/2}\rangle$ typically $\gamma_a\gg\gamma_r$, this will cause a larger decay error $\varepsilon_a$ from the dressing state.

 To address this issue we separately characterize the infidelity of gates by individually calculating two decay errors $\varepsilon_r$ and $\varepsilon_a$, as a function of $T$ in Fig.\ref{fig:temperature}(b-c). By increasing the temperature, for no-dressing cases the only Rydberg decay error $\varepsilon_r$ quickly increases closing to $\sim 10^{-2}$ at $T= 5$ mK which means the random fluctuation $\delta$ from the atomic motional dephasing will strongly affect the time-spent in the Rydberg state because no protection is performed for this state. However, with the help of dressing state $|a\rangle$ we observe that
  the $\varepsilon_r$ (here $\gamma_a=0$) obtains a big reduction and stays constant at about \textcolor{black}{$1.3\times 10^{-5}$} due to the minimization of time-spent in the Rydberg state. In addition the $\varepsilon_r$ does not vary with $T$ because $|r\rangle$ is fully protected by $|a\rangle$ in this case.
 Finally we estimate the intermediate decay error $\varepsilon_a$ from dressing state $|a\rangle$
 in Fig.\ref{fig:temperature}(c) for two dressing cases. To avoid additional errors from Rydberg state we set $\gamma_r=0$. By varying $T$ it is clear that the decay error $\varepsilon_a$ can also sustain constant owing to the perfect protection; whilst, unluckily it keeps at a quite high value $\sim 10^{-2}$ serving as the leading-order contribution to the gate infidelity (see Fig.\ref{fig:temperature}a). This is actually a result of the short lifetime of $|a\rangle$ and can be avoided by replacing with a long-lived state (see Sec. \ref{prospective}). We remark that, the robustness of $\varepsilon_r$ and $\varepsilon_a$ to {\it any} $T$ values indicates the average time-spent in both states ($|r\rangle,|a\rangle$) can be kept unvaried with the increase of $T$. Our protocol is fully immune to the Doppler-dephasing error caused by atomic thermal motion.

Accounting for the ultimate limitation by $\varepsilon_a$ in the dressing protocols, in Fig.\ref{fig:temperature}(a) we verify that, 
although the traditional no-dressing protocols can contribute a slightly higher fidelity for low temperatures $T\leq 1.5$ mK; yet they have no robustness to the Doppler errors. Remarkably, our protocol with dressing-field protection mechanism is fully robust against the Doppler errors for {\it any} temperature (at the cost of a slightly low gate fidelity $F(T)\sim 0.9906$). A prospective proposal for solving this major decay error $\varepsilon_a$ is presented in Sec. \ref{prospective}.

\subsection{Other detuning error sources}

\begin{figure}
    \centering
    \includegraphics[width=0.43\textwidth]{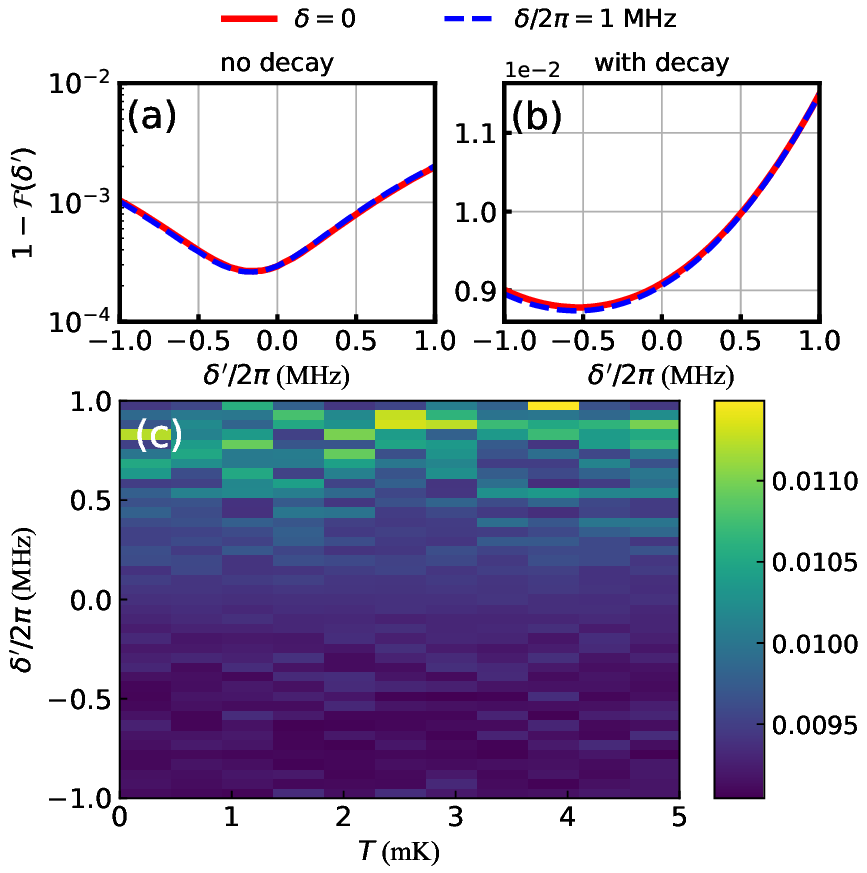}
    \caption{(a-b) The infidelity of Doppler-error immuned gates $1-F(\delta^\prime)$ as a function of other detuning error $\delta^\prime$, where the Doppler shift is $\delta=0$ and $\delta/2\pi=1.0$ MHz, for the ideal case with no decay and for the realistic case with decays, respectively. (c) The realistic gate infidelity $1-F$ as a function of both $T$ and $\delta^\prime$. Every point $(T_j,\delta^\prime_j)$ in the color plot is obtained by using an average over 300 random samplings, individually extracted from the range of an one-dimensional Gaussian distribution with $v_{rms} = \sqrt{k_BT_j/m}$ for $T_j$ and a uniform distribution of $[-\delta^\prime_j,+\delta^\prime_j]$ for $\delta_j^\prime$.}
    \label{fig:additional detuning}
\end{figure}

\begin{figure*}
\centering
\includegraphics[width=0.85\textwidth]{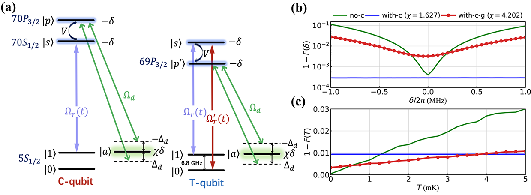} 
\caption{(a) The improved gate protocol that dresses a long-lived ground state $|a\rangle = |5S_{1/2},F=2,m_F=-1\rangle$ instead of a short-lived lower excited state. A dipole-dipole interaction $V$ drives the exchange coupling between two Rydberg pair states $|ss\rangle\rightleftarrows|pp^\prime\rangle$ (e.g.  $|70S_{1/2}S_{1/2}\rangle\rightleftarrows|70P_{3/2}69P_{3/2}\rangle$) \textcolor{black}{\cite{PhysRevA.92.042710}}. Here
states $|p\rangle$ and $|p^\prime\rangle$ ($|s\rangle$ is not protected) are directly protected through an one-photon dressing with the ground state $|a\rangle$. The Rydberg decay rates from $|s\rangle$ and $|p,p^\prime\rangle$ are $\gamma_s = 2.6$ kHz, $\gamma_{p,p^\prime} = 1.3$ kHz and the dipole-dipole interaction strength is $V=2\pi\times200$ MHz. (b-c) Gate performance based on the improved protocol in (a), denoted by {\it with-c-g} (red line with dots). For comparison the original {\it no-c} (green line) and {\it with-c} (blue line) cases are also presented. (b) The infidelity $1-F(\delta)$ as a function of the Doppler error $\delta$ in the absence of Rydberg decay. (c) The realistic gate infidelity $1-F(T)$ as a function of $T$, including the spontaneous decays from all Rydberg levels $|s\rangle,|p\rangle,|p^\prime\rangle$.}
\label{Fig.model-dipole}
\end{figure*}

It is worth pointing out that during the gate execution, other error sources such as the ac Stark shift \textcolor{black}{\cite{PhysRevA.92.022336,PhysRevA.85.042310,PhysRevResearch.5.013205}}, the fluctuation in laser frequencies \textcolor{black}{\cite{PhysRevA.101.030301,PhysRevA.107.042611,PhysRevLett.132.043201}} and so on, also serve as an unknown detuning error that leads to the realistic two-photon detuning having not only the Doppler error $\delta = -\vec{k}_r\cdot\vec{v}$ but also other unknown errors $\delta^\prime$. For a typical two-photon transition system with a large intermediate detuning a significant ac Stark shift occurs as a consequence of the uncompensated laser amplitude fluctuations \textcolor{black}{\cite{PhysRevLett.129.200501}}. According to our recent work \textcolor{black}{\cite{hou2024active}} for a modest estimation of laser amplitude deviation $\sim5.0\%$ this unknown detuning error is about $2\pi\times 0.32$ MHz.
To quantify the robustness of gate performance against other detuning errors, Fig.\ref{fig:additional detuning}(a) shows the ideal gate infidelity at $\delta=0$ and $2\pi\times1.0$ MHz by varying extra errors over a wide parameter range $\delta^\prime/2\pi\in[-1.0,1.0]$ MHz. The observed gate infidelity $1-F(\delta^\prime)$ clearly varies with $\delta^\prime$ due to the breakdown of insensitive condition. Because the realistic detuning $\chi\delta$ of dressing state $|a\rangle$ is irrelevant to $\delta^\prime$ (just relevant to $\delta$), making the robust gate no longer immune to these error sources. By increasing the values of $|\delta^\prime|$ we find the ideal gate infidelity has a same trend of exponential enhancement for different $\delta/2\pi=(0,1.0)$ MHz. In general this detuning error source contributes at the level of $<10^{-3}$ which is much smaller than the decay error $\varepsilon_a$ by more than 1 order of magnitude, and is therefore not important. This can further be verified by including the spontaneous decays. In 
Fig. \ref{fig:additional detuning}(b) we show the realistic infidelity over a range of detuning fluctuations $\delta^\prime$. As expected, the total gate infidelity sustains around $\sim 10^{-2}$ confirming trivial impact from other detuning errors, as compared to the decay error.

Finally, we consider the realistic gate infidelity in the presence of both Doppler error (induced by a finite temperature $T$) and other detuning errors $\delta
^\prime$. Figure \ref{fig:additional detuning}(c) presents the gate performance with robust pulses ($\chi=1.627$) in the ({\it with-c}) dressing case. It is clear that the infidelity numbers can always maintain around 0.01 in the presence of significant detuning imperfections or Doppler dephasing errors, which implies a powerful robustness of our gate protocol to any magnitude of errors on the two-photon detuning.

\section{Prospects for dressing with long-lived ground states} 
\label{prospective}

In Sec. \ref{realistic} we have shown the erasing of Doppler dephasing error in realizing Rydberg quantum gates, however the major weakness of which lies in a large decay error $\sim 0.01$ from the intermediate dressing state $|a\rangle$ resulting in the ultimate gate fidelity $F\approx0.9906$ even at low temperatures. In order to overcome this decay error we proceed to consider the auxiliary state to be a stable ground state so as to avoid this dominant error. In Fig.\ref{Fig.model-dipole}(a), we assume that 
the qubits are encoded into two hyperfine atomic ground states $|0\rangle$ and $|1\rangle$, and the traditional excitation to Rydberg $|s\rangle$ states is driven by a two-photon process. Whilst when two atoms are prepared in pair state $|ss\rangle$ they will experience a natural two-body F\"orster resonance by obeying $|ss\rangle\rightleftarrows|pp^{\prime}\rangle$ with a dipole-dipole interaction strength $V$ \textcolor{black}{\cite{Ravets2014,PhysRevA.94.062307,PhysRevA.96.042306}}. Here, due to the transition selection rule, we remark that two dressing fields $\Omega_d$ should drive the coupling between $|p\rangle,|p^\prime\rangle$ and $|a\rangle$ via a 297 nm ultraviolet laser \textcolor{black}{\cite{PhysRevA.99.042502}} since state $|s\rangle$ is transition-forbidden. This new choice provides $\vec{k}_a=-2\pi/\lambda_{297} =21.156 \mu m^{-1}$($\vec{k}_r$ is same) arising a larger sensitivity factor $\chi=4.202$ if $\Omega_d/\Delta_d \approx 0.463$.

The total Hamiltonian for the above improved gate protocol reads
\begin{equation}
    \mathcal{H}=\mathcal{H}_c \otimes I + I \otimes \mathcal{H}_t + V(|ss\rangle \langle pp^{\prime}|+|pp^{\prime}\rangle \langle ss|)
\end{equation}
where single-qubit Hamiltonians are
\begin{eqnarray}
    \mathcal{H}_c&=&\frac{\Omega_r(t)}{2}|1\rangle \langle s|+\frac{\Omega_d(e^{i\Delta_dt} + e^{-i\Delta_dt})}{2}|p\rangle\langle a|+\text{H.c.} \nonumber\\
    &-&\delta(|s\rangle\langle s|+|p\rangle\langle p|)+\chi\delta|a\rangle\langle a| \nonumber \\
    \mathcal{H}_t&=&\frac{\Omega_r(t)}{2}|1\rangle \langle s|+\frac{\Omega^{\prime}_r(t)}{2}|0\rangle \langle s|
    +\frac{\Omega_d(e^{i\Delta_dt} + e^{-i\Delta_dt})}{2}|p^{\prime}\rangle\langle a| \nonumber\\
    &+&\text{H.c.}-\delta(|s\rangle\langle s|+|p\rangle\langle p|)+\chi\delta|a\rangle\langle a|, \nonumber
\end{eqnarray}
 respectively. Similarly, numerical simulation adopts the Liouville-von Neumann equation: $ \partial \rho(t)/\partial t = -i[\mathcal{H},\rho]+\mathcal{L}_s[\rho]+\mathcal{L}_p[\rho]$ with two Lindblad relaxation terms 
\begin{eqnarray}
    \mathcal{L}_s[\rho] = \sum_{i\in\{0,1,a\}}(L_{is}\rho L_{is}^\dagger-\frac{1}{2}[L_{is}^\dagger L_{is}\rho + \rho L_{is}^\dagger L_{is}]) \nonumber\\
    \mathcal{L}_p[\rho] = \sum_{i\in\{0,1,a\}}\sum_{j\in\{p,p^{\prime}\}}(L_{ij}\rho L_{ij}^\dagger-\frac{1}{2}[L_{ij}^\dagger L_{ij}\rho + \rho L_{ij}^\dagger L_{ij}]) \nonumber 
\end{eqnarray}
where the Lindblad operators are
{$L_{is}=\sqrt{\gamma_s/3}|i\rangle\langle s|$, {$L_{ip}=\sqrt{\gamma_{p}/3}|i\rangle\langle p|$, $L_{ip^\prime}=\sqrt{\gamma_{p^\prime}/3}|i\rangle\langle p^\prime|$ describing the Rydberg state decays. By extending to the improved protocol that dresses with a hyperfine ground state we apply numerical optimization to both laser amplitude and (composite) phase profiles, in which the parameters are (all units are same as in Table \ref{table1})
\begin{align*}
    &\Omega_{\text{max}}=2\pi \times 8.39, \quad \omega=0.1179 \nonumber \\
    &\Omega^{\prime}_{\text{max}}=2\pi \times 7.94, \quad \omega^{\prime}=0.1287 \nonumber \\
    &\Omega_d=2\pi \times 163, \quad \Delta_d=2\pi \times 352 \nonumber \\
    &\delta_0=-2\pi \times 14.81, \quad \delta_1=2\pi \times 1.16 \nonumber\\ 
    &\delta_2=-2\pi \times 0.014, \quad T_g= 0.8
\end{align*}
We note that this new protocol benefits from a much shorter gate duration for minimizing the time-spent in Rydberg states only, more similar as the no-dressing case.

To see whether this protocol is fully robust to the Doppler dephasing error and meanwhile manifests as a high-fidelity gate, we first simulate the ideal infidelity $1-F(\delta)$ as a function of $\delta$ in Fig.\ref{Fig.model-dipole}(b). The best Doppler-error erased case with $\chi=1.627$ for Rb atoms, is comparably displayed which undoubtedly shows a full robustness to the variation of Doppler shift $\delta$. However, this new protocol has some robustness (as compared to the no-dressing case) against the change of $\delta$ values, unfortunately
it is unable to be fully robust. This can be understood by the design of two dressing fields which only drive the Rydberg $|p,p^\prime\rangle$ states achieving the protection from motional dephasing merely for these states. While other Rydberg $|s\rangle$ states can feel same magnitude of Doppler errors which are not directly protected by the presence of $|a\rangle$. Thus the resulting gate infidelity can no longer be fully immune to the Doppler dephasing error, revealing a small increase with the Doppler shift.

Next we proceed by studying the realistic gate performance $1-F(T)$ in the presence of both Doppler and Rydberg decay errors. Figure \ref{Fig.model-dipole}(c) shows the realistic gate infidelities by averaging over sufficient($N=300$) random samplings of atomic velocity for each $T$. The {\it no-c} and {\it with-c} protocols are same as in Fig.\ref{fig:temperature}(a). We note that the improved {\it with-c-g} model, although is less robust than the former {\it with-c} protocol because the infidelity slightly increase with $T$; yet the exact gate infidelity of which is mostly smaller than that of the {\it with-c} protocol over a wide parameter range of $T\in[0,4]$ mK. This is expected because the {\it with-c-g} protocol can effectively avoid the large spontaneous decay from intermediate excited states arising the gate fidelity $F\approx 0.9965$ at $T=50$ $\mu$K. Even at $T = 5.0$ mK the average gate fidelity of {\it with-c-g} can sustain at \textcolor{black}{$F\approx 0.9892$}, largely outperforming the former {\it no-c} protocol ($F\approx 0.97$). Note that for the fully-robust {\it with-c} protocol the realistic gate fidelity can stay at $F \approx 0.9906$ for {\it any} temperature.

\section{Conclusion and Outlook}

So far the ground-Rydberg dephasing error from finite atomic temperature is believed to an ultimate limitation for the observed gate fidelity, however it is rarely mitigated unless advanced technologies of atomic cooling and trapping are developed \textcolor{black}{\cite{doi:10.1126/science.aaz6801,10.21468/SciPostPhys.10.3.052,PhysRevX.8.041055}}. In the present work, we have used the protection scheme, first proposed by Finkelstein and coworkers \cite{PhysRevX.11.011008}, to demonstrate a family of high-fidelity Doppler-error erased gates, enabled by cleverly dressing with an auxiliary state that protects the Rydberg state from Doppler dephasing errors.
On the basis of robust optimal control strategy, we present several dressing protocols in alkaline Rb and Cs atoms that implement two-qubit CNOT gates using global modulations of both laser amplitude and phase profiles \textcolor{black}{\cite{PhysRevApplied.14.054062}}. All protocols are fully robust to the Doppler dephasing errors yet at the cost of a slightly large decay error from the intermediate auxiliary state, which fundamentally restricts the attainable gate fidelity of $F\approx 0.9906$ for {\it any} temperature.
In addition we also show that it is more interesting to replace with a no-lossy ground state to mitigate such incoherent spontaneous decay error achieving two-qubit CNOT gates with $0.9965$ fidelity at $T=50$ $\mu$K, although the perfect insensitivity to Doppler errors would be slightly broken due to the existence of other unprotected Rydberg states. Finally we note that robust pulses with a larger amplitude can enable significant improvements to the gate performance (see Appendix \ref{appendixb}), suggesting a promise to the gate fidelity of 0.9955 for {\it any} temperature.

This work can strongly relax the temperature constraint for achieving high-fidelity Rydberg gates by fully erasing the effect of Doppler dephasing error at {\it any} temperature, and would be worthwhile for future experimental demonstration. Before ending we have to admit that our gate protocol will be also impacted by other technical imperfections not included in the analysis \textcolor{black}{\cite{PhysRevA.72.022347,PhysRevA.101.043421}}. Whilst a gate with full robustness to certain error sources is very useful, especially to the ground-Rydberg dephasing error which plays a dominant role for high-fidelity quantum information processing in scalable neutral-atom platforms \textcolor{black}{\cite{PhysRevA.89.032334,PhysRevLett.121.123603}}. We hope such a protocol could be utilized to explore other error-tolerant mechanism in future experiments.

\begin{acknowledgments}
We acknowledge financial support from the Innovation Program for Quantum Science and Technology 2021ZD0303200; the National Key Research and Development Program of China under Grant No. 2016YFA0302001;
by the NSFC under Grants No. 12174106, No.11474094,  No.11104076 and No.11654005, by the Science and Technology Commission of Shanghai Municipality under Grant No.18ZR1412800, by the Shanghai Municipal Science and Technology Major Project under Grant No. 2019SHZDZX01 and the Shanghai talent program.
\end{acknowledgments}

\appendix

\section{Derivation of the insensitive condition}
\label{appendixa}

{\it Insensitive condition.—}We start by deriving the insensitivity condition. To ease the derivation we assume that the Rydberg state $|r\rangle$ would experience an inhomogeneous energy shift $-\delta$ due to the inevitable Doppler shift induced by the residual thermal motion of qubit atoms, where $\delta=\vec{k}_r\cdot \vec{v}$ ($\vec{k}_r$ is the two-photon wavevector and $\vec{v}$ denotes atomic velocity). The addition of an auxiliary state $|a\rangle$ leads to a new subspace $\left\{|r\rangle, |a\rangle\right\}$ based on which we introduce a coherence protection mechanism, i.e. the double-dressing scheme as proposed and experimentally verified by Finkelstein {\it et.al.}
\textcolor{black}{\cite{PhysRevX.11.011008}}. 
Here we specifically discuss the double-dressing scheme that can be applied for constructing a high-quality Doppler-error erased gate.
For an unknown shift $-\delta$ on $|r\rangle$, the auxiliary state $|a\rangle$ will obtain a similar shift $\delta^\prime = (\vec{k}_r+\vec{k}_a)\cdot \vec{v}$ originating from the same source of inhomogeneity of atomic velocity. Since we require $\delta^\prime = \chi\delta$ with $\chi=|1+\vec{k}_a/\vec{k}_r|$ a dimensionless sensitivity factor, the dressing-field wavevector $\vec{k}_a$ should be opposite to $\vec{k}_r$ meanwhile requiring a larger amplitude $|\vec{k}_a|>|\vec{k}_r|$.

\begin{figure}
\centering
\includegraphics[width=0.43\textwidth]{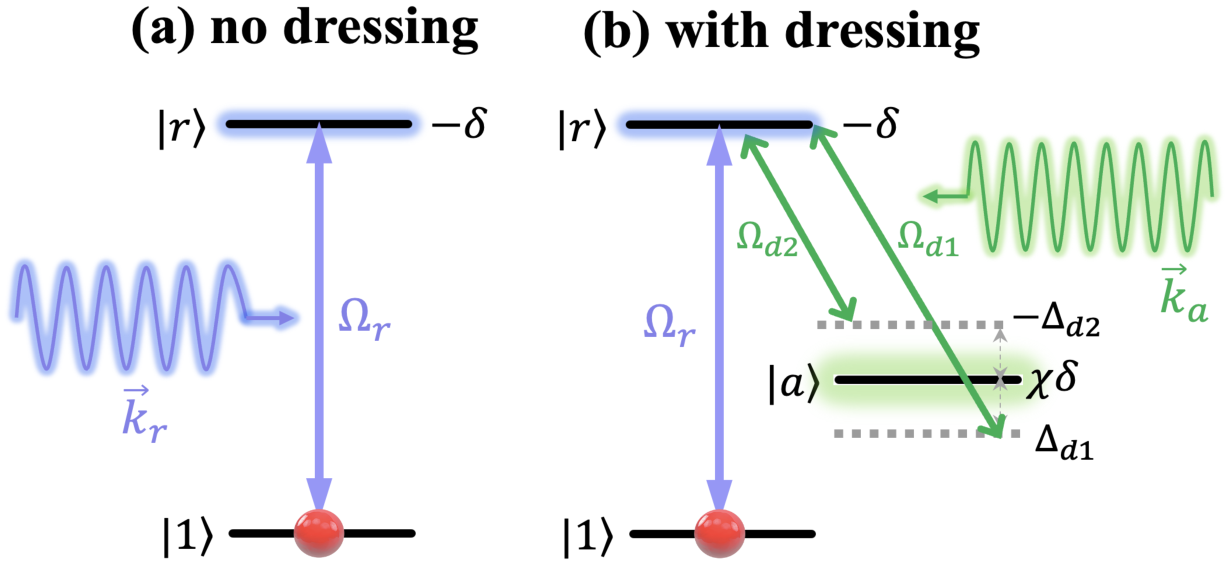} 
\caption{Coherence protection mechanism. (a) No dressing case: Single atom without dressing field is composed by the ground state $|1\rangle$ and the Rydberg state $|r\rangle$, which are coupled by a two-photon Rabi frequency $\Omega_r$. The overall transition frequency shift is denoted by $-\delta$ due to the residual thermal motion of atoms at finite temperature. (b) With dressing case: Two dressing fields with Rabi frequencies $\Omega_{d1}$ and $\Omega_{d2}$ 
and opposite detunings $\Delta_{d1}$ and $-\Delta_{d2}$, 
are applied to protect state $|r\rangle$ from motional dephasing. The auxiliary state $|a\rangle$ experiences an opposite and possibly larger transition frequency shift $\chi\delta$, where $\chi = |1+\vec{k}_a/\vec{k}_r|$ denotes a sensitivity factor.}
\label{Fig.appendixa}
\end{figure}

The double-dressing scheme as shown in Fig.\ref{Fig.appendixa}(b), involves a pair of dressing fields $\Omega_{d1}$(detuned by $\Delta_{d1}$) and $\Omega_{d2}$(detuned by $-\Delta_{d2}$). The state $|r\rangle$ is coupled to $|a\rangle$ by the dressing fields, arising the total Hamiltonian in the subspace $\left\{|r\rangle, |a\rangle\right\}$ given by $\mathcal{H}_a=\mathcal{H}_0+\mathcal{H}_1$, where
\begin{eqnarray}
    \mathcal{H}_0&=&\left[\begin{array}{cc}
    -\delta & 0 \\
    0       & \chi\delta
    \end{array}\right] \\
    \mathcal{H}_1&=&(\frac{\Omega_{d1}}{2}e^{i\Delta_{d1}t}+\frac{\Omega_{d2}}{2}e^{-i\Delta_{d2}t})\left[\begin{array}{cc}
        0 & 1 \\
        1 & 0
    \end{array}\right] 
\end{eqnarray}

To simplify the subsequent derivation, we rewrite $\mathcal{H}_0$ into a matrix form as
\begin{equation}
     \mathcal{H}_0=\frac{\delta}{2}(\left[\chi-1\right]\left[\begin{array}{cc}
    1 & 0 \\
    0 & 1
    \end{array}\right]-\left[\chi+1\right]\left[\begin{array}{cc}
    1 & 0\\
    0 & -1
    \end{array}\right])
\end{equation}
and rotate the system around the $y$ axis by using 
\begin{equation}
    U=\frac{1}{\sqrt{2}}\left[\begin{array}{cc}
    1 & -1\\
    1 & 1
    \end{array}\right]
\end{equation}
such that the rotated Hamiltonian transforming to the $\sigma_x$ representation, takes form of 
\begin{align}
&\mathcal{H}^R_a=U^{\dagger}\mathcal{H}_{a}U =U^{\dagger}(\mathcal{H}_{0}+\mathcal{H}_{1})U\nonumber\\
    &= [\frac{\Omega_{d1}}{2}(\cos{\Delta_{d1}t}+i\sin{\Delta_{d1}t})+\frac{\Omega_{d2}}{2}(\cos{\Delta_{d2}t-i\sin{\Delta_{d2}t}})] \nonumber\\
    &S_z+ \frac{\delta}{2}([\chi-1]I+[\chi+1]S_x)
\end{align}
where $I$ is the identity, and
\begin{eqnarray}
    S_x=\left[\begin{array}{cc}
    0 & 1\\
    1 & 0 \end{array}\right]\
    S_y=\left[\begin{array}{cc}
    0 & -i\\
    i & 0 \end{array}\right]
    S_z=\left[\begin{array}{cc}
    1 & 0\\
    0 & -1 \end{array}\right] \nonumber
\end{eqnarray}
Next, after introducing a unitary transformation operator for $\mathcal{H}_1$ only,
\begin{align}
    U^{'}(t)&=\mathrm{exp}(-i\int_{0}^{t}U^{\dagger}\mathcal{H}_1(t_1)U dt_1) \nonumber\\
    &=\mathrm{exp}(-iS_z[\frac{\Omega_{d1}}{2\Delta_{d1}}(\sin\Delta_{d1}t-i\cos\Delta_{d1}t+i)\nonumber\\
    &+\frac{\Omega_{d2}}{2\Delta_{d2}}(\sin\Delta_{d2}t+i\cos\Delta_{d2}t-i)])
\end{align}
we obtain the total Hamiltonian under the interaction picture
\begin{align}
    \mathcal{H}^I_{a}&=\frac{\delta}{2}[(\chi-1)I+(\chi+1)(S_+e^{(iz_1\sin\Delta_{d1}t-iz^{'}_1\cos\Delta_{d1}t-z_1)}\nonumber\\ &+S_+e^{(iz_2\sin\Delta_{d2}t+iz^{'}_2\cos\Delta_{d2}t+z_2)}+H.c.)]
    \label{interaction}
\end{align}
where $z_1=\Omega_{d1}/\Delta_{d1}$, $z^{'}_1=i\Omega_{d1}/\Delta_{d1}$, $z_2=\Omega_{d2}/\Delta_{d2}$, $z^{'}_2=i\Omega_{d2}/\Delta_{d2}$, $S_{+}$ is the Pauli matrix.

Utilizing the Jacobi-Anger identity
\begin{equation}
    e^{\pm iz\sin(\Delta_d t)}=\sum_{n=-\infty}^{+\infty}J_n(z)e^{\pm in\Delta_d t}
\end{equation}
we reduce Eq.(\ref{interaction}) to the first-order in the Magnus expansion \textcolor{black}{\cite{BLANES2009151}}
\begin{align}
    &\mathcal{H}_a^{(1)}=\frac{1}{T}\int_0^T \mathcal{H}^{I}_{a}(t_1)dt_1 \nonumber\\
    &=\frac{\delta}{2}[(\chi-1)I+J_0(z_1)J_0(z^{'}_1)(\chi+1)(S_+e^{-{z_1}}+S_-e^{z_1})\nonumber\\
    &+J_0(z_2)J_0(z^{'}_2)(\chi+1)(S_+e^{{z_2}}+S_-e^{-z_2})]\nonumber\\
    \label{magnus}
\end{align}

To find the insensitive condition we require the eigenvalues of $\mathcal{H}_a^{(1)}$ vanish (i.e. independent of $\delta$). By assuming two dressing fields with equal amplitudes and opposite detunings {\it i.e.} $\Omega_d = \Omega_{d1} = \Omega_{d2}$, $\Delta_d = |\Delta_{d1}|=|\Delta_{d2}|$, the first-order reads 
\begin{equation}
    \mathcal{H}^{(1)}_a=\frac{\delta}{2}[(\chi-1)I+2J_0(z)(\chi+1)(S_+e^{-z_1}+S_-e^{z_1})] \nonumber
\end{equation}
in which the general insensitive condition occurs at
\begin{equation}
    J_0(z_1)=\frac{(\chi-1)}{2(\chi+1)}
    \label{jocobi}
\end{equation}

{\it Numerical verification.—}To verify the validity of the insensitive condition (\ref{jocobi}), we apply it for achieving the population transfer in a single atom, in order to see whether the transfer process is perfectly robust to the variation of energy shift $\delta$. Consider a single atomic qubit comprising $|1\rangle$ and $|r\rangle$, as displayed in Fig.\ref{Fig.appendixa}(a). The atom initially in $|1\rangle$, would experience a typical state rotation $|1\rangle\to|r\rangle\to|1 \rangle$ after adopting a $2n\pi$ pulse ($\Omega_r\tau = 2n\pi$, $\tau$ is the transfer duration, $n$ is a tunable integer). 
However, since the atoms are not stationary, their inhomogeneous velocity distribution leads to Doppler dephasing which reduces the fidelity of the population transfer. In Fig. \ref{Fig.appendixb}
we numerically calculate the infidelity of state transfer by following the master equation $d\rho/dt = -i[\mathcal{H}_a,\rho]$ in the absence of any decay ($\gamma=0$), as a function of the Doppler detuning error $\delta$. Depending on the definition of $\chi = |1+\vec{k}_a/\vec{k}_r|$, without loss of generality, we adopt the sensitivity factor $\chi = (0.5,1.0,50)$ for example. Note that in the absence of dressing fields the transfer infidelity $1-F$ has an exponential enhancement with $|\delta|$ and is quickly close to 0.5 when the frequency shift $|\delta|/2\pi$ increases to 1.0 MHz. However, with the help of double-dressing protection scheme the insensitivity of the infidelity could be dramatically improved which implies that the insensitive condition can indeed make the system insensitive to the Doppler dephasing, achieving a perfect protection from the Doppler dephasing error.

\begin{figure}
\centering
\includegraphics[width=0.45\textwidth]{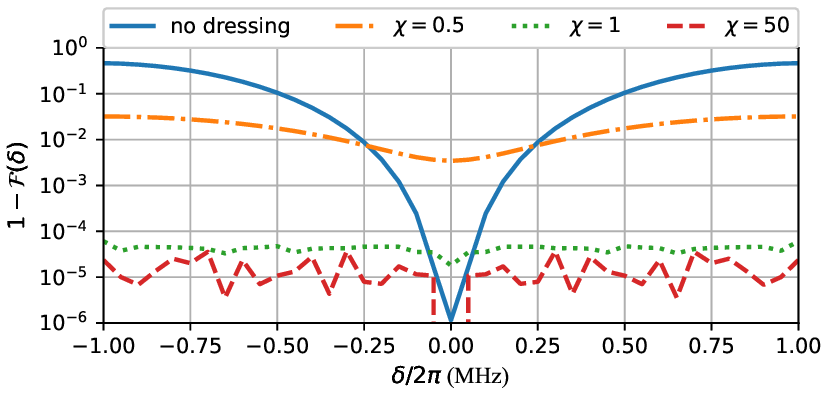} 
\caption{The infidelity $1-F(\delta)$ in the absence of any decay with different choices of $\chi=(0.5,1.0,50)$. The no dressing case is shown for comparison. All pulse parameters are found in Table \ref{tab:table1}. }
\label{Fig.appendixb}
\end{figure}

\begin{table}
\caption{\label{tab:table1}{Coefficients 
for plotting Fig.\ref{Fig.appendixb}. The transfer duration $\tau$ is fixed to 1.0 $\mu s$.}}
\begin{ruledtabular}
\setlength{\tabcolsep}{0mm}{
\begin{tabular}{ccccc}
$\quad$&\multicolumn{1}{c}{no dressing}&\multicolumn{3}{c}{with dressing}\\
\hline
($2\pi\times$MHz)&$\quad$&$\chi=0.5$&$\chi=1$&$\chi=50$\\
\hline
$\Omega_r$&1&10&3&4\\
$\Omega_d$&/&200&85&183\\
$\Delta_d$&/&45&35&460\\
\end{tabular}
}
\end{ruledtabular}
\end{table}

\begin{table*}
\caption{\label{table:appendixb}{Coefficients of the optimized laser pulses $\Omega_{\max}$, $\Omega_{\max}^{\prime}$, $\Omega_d$ and $\phi(t),\phi^{\prime}(t)$ and laser phases $\phi(t)$, $\phi^\prime(t)$ for the improved schemes ({\it with-c-improve} and {\it with-c-g-improve}) using larger laser amplitudes. The last column presents the realistic gate fidelity at $T=0$.
Other parameters are given in Table \ref{table1}.}}
\begin{ruledtabular}
\renewcommand{\arraystretch}{1.1}
\setlength{\tabcolsep}{0mm}{
\begin{tabular}{ccccccccccccccc}
Case & Auxiliary state & \multicolumn{1}{c}{$T_g$} & \multicolumn{2}{c}{$\Omega_r(t)$} & \multicolumn{2}{c}{$\Omega^{\prime}_r(t)$}  & \multicolumn{4}{c}{$\phi(t),\phi^\prime(t)$} &  $\Omega_d$ & $\Delta_d$ & $F$(realistic)\\
\hline
$\quad$ & $\quad$ & $\quad$ & $\Omega_{\max}$ & $\omega$ & $\Omega^{\prime}_{\max}$ & $\omega^{\prime}$ & $\delta_0$ & $\delta_1$ & $\delta_2$  &$\alpha$ &$\quad$ &$\quad$\\
\hline
{\it with-c-improve}  & excited state & 2.54 & 19.64 & 0.0769 & 19.29 & 0.0768 & -10.44 & 1.93 & 16.56 & 1.288 & 225.53 & 323.11 & 0.99547 \\
\hline
{\it with-c-g-improve} & ground state & 1.14 & 19.85 & 0.1179 & 19.38 & 0.1202 & 20.0 & 0.90 & -15.99 & 0.002 & 173.36 & 374.42 & 0.99998 \\
% (i) & intermediate state & 2.54 & 19.64 & 19.29 & 0.0769 & 0.0768 & -10.44 & 1.93 & 16.56 & 0.6440 & 225.53 & 0.9992 \\
% \hline
% (ii) & ground state & 3.02 & 12.18 & 13.87 & 0.1929 & 0.1654 & 20.0 & -8.46 & -16.94 & 0.6124 & 171.1 & 0.9993 \\
\end{tabular}
}
\end{ruledtabular}
\end{table*}

Moreover, we find that the exact ratio between $\Omega_d$ and $\Delta_d$ caused by a different sensitivity factor $\chi$, is quite different.
For a small $\chi = 0.5$ we have $\Omega_d/\Delta_d \approx 4.443$ and therefore the dressing state becomes near resonance there. The addition of double-dressing fields will severely affect the original state rotation between $|1\rangle$ and $|r\rangle$, arising a poor infidelity number $\sim 10^{-2}$. While for $\chi=1$, according to Eq. (\ref{jocobi}), we have $\Omega_d/\Delta_d\approx2.405$. The resulting infidelity as shown in Fig.\ref{Fig.appendixb}, stays independent of $\delta$ and achieves a very low number $< 10^{-4}$. For a larger $\chi=50$ which leads to $\Omega_d/\Delta_d\approx 0.398$, since $\Delta_d > \Omega_d$ the transfer infidelity is still perfectly protected from the fluctuation of Doppler shift, arising a lower infidelity number $\sim 10^{-5}$ as compared to the $\chi=1$ case. Therefore, an enhanced sensitivity $\chi\geq 1$
enables the use of dressing fields far-detuned and achieves a higher-fidelity state transfer, 
thus staying a high-insensitivity to the inhomogeneous Doppler shift despite the continuous exposure of the atoms to the residual thermal motion.
The potential application of this protection scheme to achieve robust Doppler-error erased gates have been profoundly discussed in the main text.

\section{Improved scheme with larger laser amplitudes}
\label{appendixb}

\begin{figure}
\centering
\includegraphics[width=0.47\textwidth]{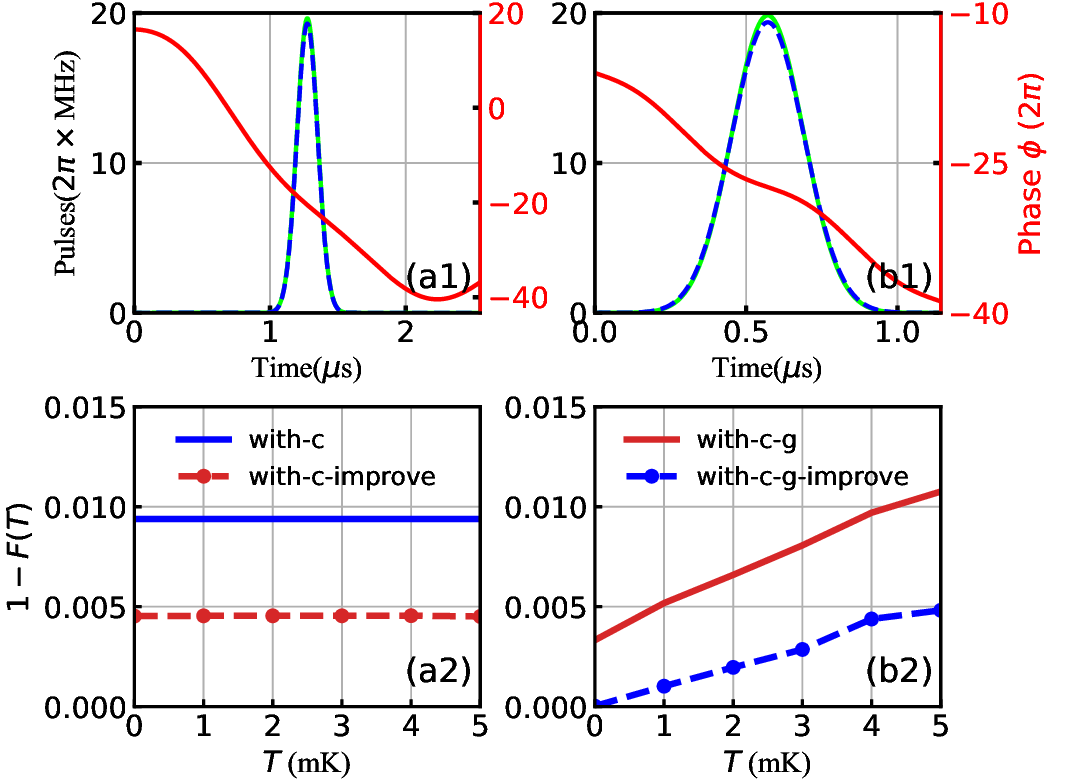} 
\caption{Gate performance by using larger laser amplitudes. (a1-b1) Optimized profiles for laser amplitudes and phases, corresponding to the intermediate excited-state dressing and ground-state dressing protocols, respectively. All linetypes are same as in Fig.2(b). (a2-b2) The realistic gate infidelity $1-F(T)$ as a function of $T$. For comparison the original cases ({\it with-c} and {\it with-c-g}) are displayed by solid lines. Two improved cases ({\it with-c-improve} and {\it with-c-g-improve}) with larger amplitudes are denoted by dash-dotted lines.
All relevant gate parameters are presented in Table
 \ref{table:appendixb}. 
}
\label{Fig.appb}
\end{figure}

In the main text we have presented the realization of Doppler-error insensitive gates via different dressing protocols. However the exact gate fidelity number is still not satisfactory. In this section we discuss the realistic gate performance using a higher two-photon Rabi frequency by extending the restriction to $(\Omega_{\max},\Omega_{\max}^\prime)/2\pi \leq 20$ MHz, possibly done by ultrafast pulsed-laser technique \textcolor{black}{\cite{PhysRevLett.114.203002,Chew2022,doi:10.1126/science.aau1949}}, and show that both the {\it with-c} and {\it with-c-g} protocols can further be improved by substantially decreasing the time-spent in intermediate excited and Rydberg states.

To perform a deep optimization, we additionally introduce a slightly more general ansatz for the phase profile, given by
\begin{equation}
    \phi(t) = \delta_0 t + \delta_1\sin(\frac{4\pi t}{T_g}) + \delta_2\cos(\frac{\alpha\pi t}{T_g})
\end{equation}
where coefficient $\alpha$ is added for further fine-tuning, and a smoothly-modulated Gaussian pulse for laser amplitude is kept.
Inspired by the results in the main text, we note that the {\it with-c} protocol tends to require the laser pulse having a narrowed peak
with a long tail for minimizing the intermediate-state population, e.g. for $\chi=1.627$, the total time-spent in the Rydberg and auxiliary states is $(P_r,P_a)\approx(0.0884,0.0554)$ $\mu$s approximately. In contrast, the {\it with-c-g} protocol is driven by using a slightly global pulse within a shorter gate time (more similar to the no-dressing case), which arises
$(P_p,P_s,P_a)\approx(1.15\times10^{-4},0.1691,6.39\times 10^{-5})$ $\mu$s when $\chi = 4.202$. We emphasize these time-spent numbers are obtained for the cases with spontaneous decays. Here, we re-optimize all gate parameters for two protocols ({\it with-c} and {\it with-c-g}) by applying a higher laser amplitude, denoted as {\it with-c-improve} and {\it with-c-g-improve}, respectively.

In Fig.\ref{Fig.appb}(a1-b2), the optimized pulse amplitude and phase profiles as well as the realistic gate infidelity as a function of $T$ are shown. The corresponding gate parameters are summarized in Table \ref{table:appendixb}. With higher laser amplitudes we clearly find a significantly improved gate quality, greatly outperforming the {\it with-c} and {\it with-c-g} protocols. The {\it with-c-improve} protocol shows a gate infidelity of $1-F\approx0.0045$ for {\it any} temperature, reducing the infidelity number by $52.1\%$ as compared to the {\it with-c} case. As turning to the {\it with-c-g-improve} scheme we interestingly find that at $T=50$ $\mu$K the realistic gate fidelity can retain as high as $F\approx 0.99991$, and even at $T = 5$ mK the gate fidelity preserves $F\approx 0.9955$. These improvement can be attributed to the fact of a strong time-spent minimization in the intermediate lossy states, leading to $(P_r,P_a)\approx (0.0264,0.0209)$ $\mu$s for the 
{\it with-c-improve} protocol and $(P_p,P_s,P_a)\approx (5.53\times 10^{-4},0.1397,1.88\times 10^{-4})$ $\mu$s (here the unprotected $|s\rangle$ is dominant) for the 
{\it with-c-g-improve} protocol.

 To conclude, these improved protocols with larger laser amplitudes hold great promises for higher-quality Rydberg quantum gates. The fully robust {\it with-c-improve} protocol can fundamentally relax the temperature requirement for quantum gate operations. While the strongly robust {\it with-c-g-improve} protocol is able to achieve a super-high gate fidelity at colder temperatures, possibly reaching a new milestone for two-qubit quantum gates in the neutral-atom computing platform. Both of them deserve an experimental demonstration in the near future.


% \bibliography{apssamp}% Produces the bibliography via BibTeX.

\end{document}